\documentstyle [preprint,aps] {revtex}
\begin{document}
\draft

\title{Pion-nucleus elastic scattering on $^{12}$C, 
$^{40}$Ca, $^{90}$Zr, and $^{208}$Pb \break at $400$ and $500$ MeV}

\author{George Kahrimanis,$^{1,\alpha}$ George Burleson,$^{2}$
C.~M.~Chen,$^{4}$, B.~C.~Clark$^{4}$,
Kalvir Dhuga,$^{5}$ D.~J.~Ernst,$^{6}$ J.~A.~Faucett,$^{2,\beta}$
H.~T.~Fortune,$^{7}$ S.~Hama,$^{8}$
Ahmed Hussein,$^{9}$
M.~F.~Jiang,$^{6}$
K.~W.~Johnson,$^{1}$ L.~Kurth Kerr$^{4}$, Scott Mathews,$^{5}$
John McGill,$^{10,\gamma}$ C. Fred Moore,$^{1}$ 
Shaul Mordechai,$^{11}$ 
Christopher L.~Morris,$^{10}$ John O'Donnell,$^{7,\beta}$
Mike Snell,$^{1}$ Mohini Rawool-Sullivan,$^{10}$ L.~Ray,$^{1}$
Charles Whitley,$^{1,\beta}$ and Allen L. Williams$^{1,\delta}$}

\address{$^1$ Department of Physics, University of Texas at Austin,
Austin, Texas 78712-1081}
\address{$^2$ Department of Physics, New Mexico State University, Las
Cruces, New Mexico 88003}
\address{$^3$ Saint John's and Saint Mary's Institute of
Technology, \#499 Sec. IV Tam-King Rd., Tam-Sui, Taiwan 23135}
\address{$^4$ Department of Physics, The Ohio State University, Columbus,
Ohio 43210}
\address{$^5$ Department of Physics, George Washington University, 
Washington, DC  20052}
\address{$^6$ Department of Physics and Astronomy, Vanderbilt
University, Nashville, Tennessee 37235}
\address{$^7$ Department of Physics and Astronomy, University of
Pennsylvania, Philadelphia, Pennsylvania 19104}
\address{$^8$ Hiroshima University of Economics,
Hiroshima 731-01, Japan}
\address{$^9$ University of Northern British Columbia, Prince George, 
British Columbia, Canada V2N 4Z9}
\address{$^{10}$ Los Alamos National Laboratory, Los Alamos, New Mexico, 
87545}
\address{$^{11}$ Ben-Gurion University of the Negev, Beer-Sheva 84105, 
Israel}
\address{$\alpha$ Current address: 2808 35$^{th}$ Street \#5E, Long Island 
City, New York 11103}
\address{$\beta$ Currently at Los Alamos National Laboratory}
\address{$\gamma$ Currently at the University of Texas at Austin}
\address{$\delta$ Currently at the Johns Hopkins Oncology Center, Division of
Radiation Oncology, 600 North Wolf Street, Baltimore, Maryland 21287-8922}

\maketitle

\newpage

\begin{abstract}
Pion-nucleus elastic scattering at
energies above the $\Delta(1232)$ resonance is 
studied using both $\pi^+$ and $\pi^-$ beams on $^{12}$C,
$^{40}$Ca, $^{90}$Zr, and $^{208}$Pb. The present data provide an opportunity
to study the interaction of pions with nuclei at energies 
where second-order corrections to impulse approximation calculations
should be small. 
The results are compared with other data sets at similar energies, and with
four different first-order impulse approximation calculations.
Significant disagreement exists between the calculations and
the data from this experiment.
\end{abstract}


\section{INTRODUCTION}

The study of $\pi^{\pm}$-nucleus elastic differential 
scattering cross sections with incident pion 
energy above 300 MeV, {\it i.e.}, beyond the range of the first 
pion-nucleon resonance, $\Delta(1232)$, has been motivated by several 
considerations.  Because the wavelength of the pion is shorter at 
higher energies it can be a 
sensitive probe of spatial distributions in a nucleus.
The pion-nucleon two-body interaction becomes weaker above
resonance allowing the pion to penetrate deeper into the nucleus.
The two-body total 
cross section drops to 25 mb (isospin-averaged) at 500 MeV, about $18\%$ of 
its value on the peak of the resonance. 
In addition, the smaller two-body cross section implies that second-order 
corrections to first-order multiple-scattering optical potentials 
should be small.  Thus, significant differences between 
the measured elastic cross sections and the first-order impulse 
approximation (IA) calculations, especially at forward angles, 
could signal the failure of the IA which would pose a number of 
interesting physics questions. 

At energies above the $\Delta$, pion elastic differential cross sections 
at 800 MeV/c were measured at BNL for $^{12}$C and 
$^{40}$Ca~\cite{marl84}.  The data were 
in qualitative agreement with a number of different
calculations, however the differential cross sections at forward angles 
were underpredicted by all of 
them~\cite{clark85,ernst93,chen93,john92,arim1,arim2}.  
The 15\% normalization error in the data, as well as some uncertainties
in the input to the calculations, caused difficulty in pin-pointing
the reasons for the discrepancy. More recent pion-nucleus data at 400 MeV 
for $^{28}$Si from LAMPF~\cite{rawo94} and for $^{12}$C at 610, 710,
790, and 895 MeV/c and $^{208}$Pb at 790 MeV/c from 
KEK~\cite{tak-diss,kek95,tak-proc} are now available.  The
overall agreement between theory and 
experiment is qualitative at best for $^{12}$C but is 
quite reasonable for $^{208}$Pb.  As will be discussed below, the 
discrepancy at forward angles is less pronounced for these data than for 
the BNL data or the present experiment.

In Sec. II the experiment is described. Section III briefly discusses
the four theoretical calculations which are compared with the present
experimental data. Comparisons with previous data are also given.
A summary is given in Sec. IV.

\section{EXPERIMENT}

In this work we present elastic scattering data for
$\pi^\pm$ on $^{12}$C, $^{40}$Ca, $^{90}$Zr, 
and $^{208}$Pb at pion laboratory kinetic energies of 400 and 500 MeV.
The data were taken at the Los Alamos Meson Physics Facility. 
At 400 MeV one is still on
the high-energy tail of the $\Delta$. 
At 500 MeV, the pion-nucleon 
center-of-mass total energy is 1451 MeV 
which is near the P$_{11}$ (1440), the next
resonance above the $\Delta(1232)$. 

Data were obtained using the P$^3$ east channel at the Clinton P.\
Anderson Meson Physics Facility 
and the Large Acceptance Spectrometer (LAS). The P$^{3}$ channel
provides a pion beam, produced by interactions of the $800$~MeV primary
proton beam in
the $^{12}$C production target. In the original design the pion beam
was achromatic at the experimental 
target, {\it i.e.}, the mean momentum of the 
incident pions would not be correlated with 
position. For the purposes of this 
work a dispersed beam tune was developed in order to improve
the resolution in the missing mass spectra.
With this tune the momentum dispersion at the target was $0.5\%/cm$.  
The angle of incidence was also correlated with position at the target.
The energy spread of the incident pion beam was about $10$ MeV at the
$400$ MeV setting and about $12$ MeV at the $500$ MeV setting.

The outgoing pions were detected using a modified version 
of the Argonne Large Acceptance Spectrometer, a Q-Q-D system. Four drift
chamber planes, measuring both horizontal and vertical positions and angles,
were located between the second quadrupole magnet and the dipole magnet.  
Four wire chamber planes followed by two sets of scintillators 
were located after the analyzing dipole magnet. 
All of the wire chambers were delay-line-readout drift chambers\cite{Atencio81}.
In order to minimize Coulomb multiple scattering and improve the momentum
resolution 
helium bags were installed along the entire six-meter distance from the target 
to the rear wire chambers, the target was enclosed in a vacuum chamber, and 
thin (6 $\mu$m mylar) windows were used on the wire chambers.  The apparatus
included an 
additional sweep magnet between the target and the spectrometer  
which was adjusted to bend the paths of particles with the nominal momentum 
of the spectrometer setting by $10^{\circ}$ in the horizontal 
plane, and to steer particles of the opposite charge away from the 
spectrometer. The acceptance of the spectrometer was $\pm5^{\circ}$ in
scattering angle (horizontally), $\pm2.5^{\circ}$ in the vertical
direction, and $\pm10\%$ in $\Delta p/p$.

Three beam monitors were used, each supplying a relative measurement of the
integrated current
for every run.  A toroidal counter, located just 
downstream of the production target, monitored the proton beam.  
An ion chamber was located in the production target box.  
Another ion chamber was placed at the exit of the beam pipe.  
With any fixed beam tune the ratio of any 
two beam-monitor readouts was constant within $1\%$.  The hardware trigger 
consisted of a coincidence between any front-chamber signals 
with the two scintillators at the end of the spectrometer.   
The target thicknesses were 290~mg/cm$^{2}$ for 
CH$_2$, 541~mg/cm$^{2}$ for $^{12}$C, 400~mg/cm$^{2}$ for $^{40}$Ca, 
672~mg/cm$^{2}$ for $^{90}$Zr and 421~mg/cm$^{2}$
for $^{208}$Pb. 
The nominal target thickness error is 3\% for each target.
The $^{90}$Zr and $^{208}$Pb targets were enriched in isotopic purity to
greater than 95\% and the C and Ca targets were of natural isotopic abundance.


The scattered pion momentum, the coordinates and angles at the target
location were calculated using Taylor series expansions about the central
trajectory through the spectrometer. In addition, redundancy checks (we
measured four positions and four angles to describe trajectories given by
two positions, two angles and one momentum) were calculated as differences
between the measured rear angles and rear angles calculated using all
coordinate information excluding the rear angle measurements.  The
polynomial coefficients for the target positions were calibrated by using
horizontal and vertical rod targets.  Coefficients for the angle
calculations were obtained by using the beam at reduced intensity and
changing the angle by moving the spectrometer around zero degrees. The
coefficients for the momentum calculation were obtained by scanning an
elastic peak across the focal plane of the spectrometer.

The beam momentum--position correlation at the target was directly
measured using data taken with a heavy target, $^{208}$Pb, for 
which finite angular resolution effects on the missing mass spectra are
small since the recoil energies are insignificant compared to the
scattered pion kinetic energy.  The resulting  width for the elastic peak in the
missing mass spectra of the heavier targets was as small as 1.7~MeV for
beam energies of 400~MeV.

The calculation of missing mass include the recoil energy with scattering
angle, measured in the spectrometer. The effect of a correlation in the beam
direction with target position on the scattering angle was included in this
calculation. These effects are important for scattering from hydrogen
because of the large variation in kinetic energy with scattering angle 
of more than $1$~MeV/deg at large angles. 
The finite angular resolution, due to both angular dispersion in the beam
and the resolution of the angle determination in the spectrometer, limited
the missing mass resolution for hydrogen.
By comparing the width of the $^{208}$Pb elastic peak with 
the corresponding elastic peak for hydrogen at the same scattering angle
we obtained a 
direct measure of our angular resolution, which was 
${1.2^{\circ}}$ FWHM.

The measurement of position and angles both before and after the
spectrometer dipole 
provided a useful capability for rejecting
particles that underwent collisions or decay downstream from the target. The
rear angles were calculated as a polynomial of front positions, front
angles, and momentum.   The difference between the calculated and measured
angle was typically about 2 mrad. Good events were selected with a cut on
this quantity of $\pm$5 mrad.

A combination of pulse-height and time-of-flight information from the
scintillators was used to reject proton events.
Figure~\ref{pid} shows the separation of
proton events from those of lighter particles, in a
two-dimensional histogram of the time of flight {\it vs.} the geometric mean
of the heights of the pulses coming from the two scintillators.
This technique was not able to distinguish pions from electrons and muons.
Electrons were eliminated by using a Cherenkov counter in the focal plane
and muons due to pion decays in the spectrometer were eliminated by the
angle checks.  The fraction of events due to muons elastically scattered in the
target is estimated to be smaller than 0.001~\cite{Wer75}.

Figure~\ref{peaks} shows typical missing-mass spectra at $T_{\pi}=400$
MeV and $\theta_{lab}=33^{\circ}$ (averaged over $8^{\circ}$).
Typically there are several peaks in each missing-mass histogram,
corresponding to the ground state and several excited states of the 
target nucleus. We fitted each histogram with a Gaussian peak shape.  
The overlap between the ground-state peak and the inelastic peaks was 
usually small, since the resolution was $1.7$ MeV FWHM at $T_{\pi}=400$ 
MeV, and about $1.9$ MeV FWHM at $500$ MeV.

Each set of data was partitioned into $0.5^{\circ}$ bins in the 
angle of scattering (in lab coordinates). In each experimental run 
({\it i.e.} for each spectrometer setting, target, incoming energy, and pion 
polarity) we recorded data that spanned 17 bins. We incremented the 
angle of the spectrometer settings by $6^{\circ}$, so that any two 
adjacent settings had five bins in common.

The uncertainty in the absolute scattering angle was estimated to be
$\pm 0.85$~deg which arises from the uncertainty in the absolute zero degree
scattering position ($\pm 0.60$~deg) and the locations of the survey marks
on the floor ($\pm 0.25$~deg) for the spectrometer angle settings.  The
absolute scattering angles were determined (to $\pm 0.60$~deg) by placing
the spectrometer in the beam and by measuring the energy difference
between the elastic peaks for pion scattering from $^1$H and $^{208}$Pb
and using kinematics to determine the scattering angle.

The normalization 
involved two stages. First, we found each bin's relative 
normalization (referred to as the relative component of acceptance)
with respect to all other bins for the same beam tune and target
dimensions (the height of the $^{90}$Zr target was smaller than the rest of the
targets, so we used a separate normalization for those runs). This relative
normalization was fixed for given target dimensions and beam tune, 
regardless of the spectrometer angle. 
Then, for each beam tune and 
target shape we determined an overall normalization such that
$\pi - p$ scattering data from CH$_{2}$ matched the 
Arndt partial-wave fit to the corresponding 
pion-nucleon cross section~\cite{arndt}.

The relative component of the acceptance of each run was treated 
as a function of two variables, the momentum of the particle 
(more precisely, the fractional deviation from the nominal setting of 
the spectrometer) and the bin number.  With fixed beam tune and 
target dimensions, this function was fixed, 
independent of target material or spectrometer angle. The dependence of 
the relative component of acceptance of each bin on momentum was 
found by varying the momentum setting of the spectrometer and comparing 
yields. The dependence on the bin number was found by comparing yields 
at the same angle in several overlapping runs.  The latter method 
was complemented by a trial adjustment of the 17 relative bin acceptances 
so that the resulting cross sections smoothly matched as well as possible in 
$\chi^2$ to a high order polynomial (order 15 
to 19 worked best).  The fitting involved data from several targets 
and resulted in no significant modification of the previous acceptance 
coefficients. The relative normalization procedure proved quite robust when the 
statistics were adequate.

The overall normalization uncertainty in this experiment is $16\%$. This error
arises from several components: $3\%$ in the target thickness (which is counted
twice, since it applies to the normalization targets as well), $4\%$ in each
overall normalization coefficient, $1\%$ in the beam count, $1\%$ in the
isotopic purity, $1\%$ in each bin's relative normalization coefficient, an
estimated $10\%$ arising from the systematic error we describe next, and an
uncertainty of $11\%$, at most, in determining the efficiency of the detection 
setup.

The only significant systematic error in this
experiment arises from treating the relative component of the acceptance of
each bin as a function of two variables only, while there was indication of a
small dependence
upon at least another variable; that might be, for instance, a target
coordinate. But our statistical sample was inadequate for taking that effect
into account. Instead, we have estimated its contribution to the overall
uncertainty.

\section{Discussion and Analysis}

The present experimental elastic differential cross sections are compared 
with four different theoretical first-order IA calculations.
All have been used previously to investigate pion and kaon nucleus
elastic scattering. The first model discussed is based on the relativistic 
impulse approximation (RIA) which has been used with success in describing 
proton-nucleus scattering~\cite{cl83,lanny}.  For the scattering 
of spin zero mesons the 
Kemmer-Duffin-Petiau (KDP) equation~\cite{kemm,duff,pet} rather than 
the Dirac equation is used.  This KDP-RIA approach has been used to 
obtain optical potentials for both kaon-nucleus and pion-nucleus elastic 
scattering~\cite{clark85,ku94,mi96,ku95}.  The construction of 
the Lorentz scalar and vector optical potentials parallels that of 
the usual RIA.  The two-body input consists of scalar and vector neutron 
and proton mean-field Hartree densities~\cite{furn87} and the 
empirical meson-nucleon Arndt amplitudes~\cite{arndt}.
 
These calculations are shown by solid lines in 
Figs.~\ref{george_4c} through~\ref{george_4pb}.  As in the Dirac case, 
the scalar and vector potentials are large 
and tend to cancel. These optical potentials, along with the Coulomb 
potential obtained from the empirical charge distribution from electron 
scattering, are used in the second-order KDP wave equation which is solved 
by partial wave analysis to get the elastic scattering differential
cross sections. A non-relativistic impulse approximation, NRIA,
optical potential is also obtained using the same Arndt
amplitudes and the same vector neutron and proton mean-field
Hartree densities, see Ref.~\cite{ku94}. These results are given by
the dotted lines in 
Figs.~\ref{george_4c} through~\ref{george_4pb}.  These 
optical potentials, along with the Coulomb potential, are used in 
the Schr\"odinger equation and the observables 
are obtained by partial wave analysis. Both calculations include the 
effect of folding over the $1.20^{\circ}$ FWHM angular resolution uncertainty. 

As mentioned above, although the scalar and vector potentials are very 
large, in some cases over 1500 MeV at the origin, the resulting effective 
central potentials are modest in size.  In Fig.~\ref{pbcen} we 
show the KDP-RIA real and imaginary potentials 
for Pb$(\pi^{\pm},\pi^{\pm})$ at 400, 500 and 662.6 MeV. The NRIA 
potentials are also shown for 500 MeV.  The imaginary central 
potentials are remarkably energy independent. This is true for 
all of the targets considered in this work.  The values of the volume 
integral per nucleon, $J/A$,  are between $-$240 and $-$280 MeV-fm$^{3}$ 
for all targets and energies.  The real central potentials exhibit more 
energy dependence but they are always repulsive in 
this energy range\cite{fnote}.  

It is of interest to compare the two models above with results from the 
more complete momentum-space multiple-scattering 
optical potential, ROMPIN, developed by Chen, Ernst, and 
Johnson~\cite{ernst93,chen93,john92}.  These authors have obtained 
a first-order optical potential which uses covariant kinematics, 
phase-space factors and normalizations, off-shell two-body amplitudes, 
and exact Fermi-averaging integrals.  The momentum-space optical potential 
is used in a Klein-Gordon equation and the differential cross 
sections are obtained.  The results of the ROMPIN calculations are 
shown by dashed lines in Figs.~\ref{george_4c} through~\ref{george_zr}.  
In addition, the authors of Ref.~\cite{ernst93} have developed an eikonal 
model for comparison with their ``model-exact" calculation. The 
eikonal calculations are given by the dot-dashed curves in 
Figs.~\ref{george_4c} through~\ref{george_4pb}.  Both the 
momentum-space optical and eikonal models use the 
same on-shell pion amplitudes~\cite{arndt} and the same Hartree-Fock wave
functions~\cite{ng69,be75} which have been corrected for center-of-mass
motion. Both calculations have been folded over the $1.2^{\circ}$ FWHM 
angular resolution uncertainty.  

The first observation is that all of these four calculations agree with 
each other to within 10\%-15\% at small angles.  The noticeable 
differences at larger angles could reasonably be attributed 
to differences in the nuclear structure input, the differences in 
the construction of the optical potentials, and the different one-body 
equations used.  Other explanations are, of course, possible.
The calculations generally agree better with each other than 
they do with the data. This is especially true as the target mass 
increases.  The difference between the measured small angle 
differential cross sections and the calculated values 
is pronounced for all targets at both energies, though it is difficult 
to see this clearly on the log-scale of the figures~\cite{fnote}.  In 
addition, there is a systematic difference between theory and experiment
regarding the position of the first diffraction
minima. In every case the first minima for both the KDP-RIA and the
ROMPIN calculations occur at angles which are $1.0^{\circ}$ 
to $2.0^{\circ}$ larger than the corresponding minima in the data.  In 
a similar analysis of the 400 MeV $^{28}$Si$(\pi^{\pm},\pi^{\pm})$ 
LAMPF data, the position of the diffraction dip is in much better agreement 
with experiment than evidenced here for any of the targets~\cite{ku94}.   
Both of these features will be discussed again when comparisons 
with other experimental data are given.

In Fig.~\ref{george_qdat} the KEK, BNL, and LAMPF $^{12}$C$(\pi^-,\pi^-)$ 
differential cross sections are plotted as a function of the momentum 
transfer $q$.  Because all four of the calculations described above agree 
with each other at forward angles, {\it i.e.} up to the first diffraction 
minima, only the KDP-RIA calculations are shown.  The calculations agree 
with the forward angle cross sections to within the normalization errors 
except for the 800 MeV/c BNL data and the 400 and 500 MeV LAMPF data.  In 
every case, except for the 710 MeV/c KEK data, the diffraction minima in 
the KDP-RIA calculations occur at angles about $1.5^{\circ}$ to 
$2.0^{\circ}$ larger than the minima in the data.  In these calculations 
recoil is not taken into account, however the ROMPIN calculations do 
address the question of recoil and, as is shown in Fig.~\ref{george_4c}, 
the ROMPIN and KDP-RIA dip positions agree to within $0.5^{\circ}$, 
so the reason for the discrepancy is not clear.  The 
results for $^{12}$C$(\pi^+,\pi^+)$ at 400 and 500 MeV and 
800 MeV/c shown in Fig.~\ref{cpip_bcc} also indicate 
a similar result regarding the position of the diffraction minima.  
The calculated small angle cross sections for 800 MeV/c underpredict the 
BNL data but overpredict the LAMPF data.  

Figures~\ref{capip_bcc} and~\ref{capim_bcc} compare 
the $^{40}$Ca$(\pi^{\pm},\pi^{\pm})$ BNL results with those of 
the present experiment.  The underprediction of the small angle BNL 
data is evident.  However, the positions of the diffraction minima are 
well reproduced by the calculations, in disagreement with the results 
for the current experiment where the minima are shifted.  Of course, 
it is possible to fit the 400 and 500 MeV data by a simple two-parameter 
scaling of the potentials, but this does not give much insight 
into the origin of the discrepancy.  
Figures~\ref{pbpip_bcc} and~\ref{pbpim_bcc} 
show the results for $^{208}$Pb$(\pi^{\pm},\pi^{\pm})$.  The higher energy 
data at 790 MeV/c (662.7 MeV) are from KEK~\cite{tak-diss,tak-proc}.  
The disagreement between the present experiment and the calculations is 
profound while agreement with the KEK data is reasonable.  
The 10\% uncertainty in the elementary amplitudes does not result in 
calculated cross sections which resolve the discrepancy~\cite{fnote}.  
Using more recent determinations of the 
relativistic mean-field Hartree densities, which incorporate nonlinear 
chiral symmetry and broken scale invariance~\cite{fts}, do not 
change the results~\cite{fnote}.  Referring to the central 
potentials shown in Fig.~\ref{pbcen} does not reveal any anomalous 
behaviour with energy.
It is possible that some of this discrepancy may be accounted for by the
absolute angle uncertainty (0.85$^{\circ}$) in the data, but not all of it.
At this time a full understanding of
the source of the diffraction minima angle discrepancy is 
not available. 

\section{Summary}

Pion-nucleus elastic scattering at
energies above the $\Delta(1232)$ resonance has been
studied using both $\pi^+$ and
$\pi^-$ beams on $^{12}$C,
$^{40}$Ca, $^{90}$Zr, and $^{208}$Pb targets.
These new data were compared with other data sets at similar energies and with
four theoretical calculations.
Certain aspects of the predictions are in
qualitative agreement with the data; however
significant disagreements are seen, particularly in the angular positions of the
diffractive minima, which are generally not evidenced in comparisons with
other experiments. This is particularly true for 
$^{90}$Zr and $^{208}$Pb.  The cause of these discrepancies 
has not yet been identified, and may wait until other experiments 
using heavy targets become available in this energy range.

\acknowledgments

This work was supported in part by the United States Department of
Energy, the Robert A.\ Welch Foundation, the U.S.--Israel Binational
Science Foundation, and the National Science Foundation.

\begin{figure}
\caption{2-d histogram of time of flight {\it vs} the geometric mean of 
the pulse heights from the two scintillators, $S2$ and $S3$. Events in the 
upper group are identified as protons, and those inside the selected 
box as pions or lighter particles.  The latter are few, and are expected to 
be almost totally rejected by further kinematic tests. The data were taken 
with $^{208}$Pb, the central scattering angle set at 12$^{\circ}$, and 
$T_\pi=500$ MeV. The scales are linear and are in arbitrary units.}
\label{pid}
\end{figure} 
\begin{figure}
\caption{Un-normlized missing-mass spectra of $\pi^-$ scattering at 
incoming energy 400 MeV, central scattering angle set at 33$^{\circ}$, 
from $^{12}$C, $^{40}$Ca, $^{90}$Zr, and $^{208}$Pb. The bin size is 
$100$ keV. The number of events has not been folded with acceptance, which 
varies with the missing mass.  In the cases of $^{90}$Zr and $^{208}$Pb, 
there are excited states between the $2^+$ and $3^-$ states.}\label{peaks}
\end{figure} 
\begin{figure}
\caption {The differential cross sections for $^{12}$C$(\pi^{\pm},\pi^{\pm})$ 
at 400 MeV (scaled by $10^{-1}$) and 500 MeV (scaled by $10^{+1}$).  The 
solid curves are from KDP-RIA calculations, the dashed curves are from 
ROMPIN calculations, the dotdash curves are from eikonal calculations, and 
the dotted curves are from NRIA calculations.} \label{george_4c}
\end{figure}
\begin{figure}
\caption {The differential cross sections for $^{40}$Ca$(\pi^{\pm},\pi^{\pm})$ 
at 400 MeV (scaled by $10^{-1}$) and 500 MeV (scaled by $10^{+1}$). The
solid curves are from KDP-RIA calculations, the dashed curves are from
ROMPIN calculations, the dotdash curves are from eikonal calculations, and
the dotted curves are from NRIA calculations.} \label{george_4ca}
\end{figure}
\begin{figure}
\caption {The differential cross sections for $^{90}$Zr$(\pi^{+},\pi^{+})$ 
at 400 and 500 MeV and $^{90}$Zr$(\pi^{-},\pi^{-})$ at 400 MeV.  The
solid curves are from KDP-RIA calculations, the dashed curves are from
ROMPIN calculations, the dotdash curves are from eikonal calculations, and
the dotted curves are from NRIA calculations.} \label{george_zr}
\end{figure}
\begin{figure}
\caption {The differential cross sections for 
$^{208}$Pb$(\pi^{\pm},\pi^{\pm})$ at 400 MeV (scaled by $10^{-1}$) 
and 500 MeV (scaled by $10^{+1}$).  The solid curves are from KDP-RIA 
calculations, the dotdash curves are from eikonal calculations, and 
the dotted curves are from NRIA calculations.  ROMPIN calculations were 
not available for this target.} \label{george_4pb}
\end{figure}
\begin{figure}
\caption {The real and imaginary effective central potentials 
for $^{208}$Pb$(\pi^+,\pi^+)$ and $^{208}$Pb$(\pi^-,\pi^-)$.  
Solid curves are for KDP-RIA at 400 MeV, dashed are 
KDP-RIA at 500 MeV, dotdash are NRIA at 500 MeV and 
dotted are KDP-RIA at 662.6 MeV (790 MeV/c). }\label{pbcen}
\end{figure}
\begin{figure}
\caption {The KEK, BNL, and LAMPF $^{12}$C$(\pi^-,\pi^-)$ differential 
cross sections as a function of the momentum transfer $q$.  Solid curves 
are from KDP-RIA calculations.  (A) shows LAMPF data at 400 MeV scaled by 
$10^{+7}$, (B) shows KEK data at 486.2 MeV (610 MeV/c) scaled by 
$10^{+5}$, (C) shows LAMPF data at 500 MeV scaled by $10^{+3}$, (D) shows  
KEK data at 584.02 MeV (710 MeV/c) scaled by $10^{+1}$, (E) shows KEK 
data at 662.7 MeV (790 MeV/c) scaled by $10^{-1}$, (F) shows BNL data at 
672.5 MeV (800 MeV/c) scaled by $10^{-3}$, and (G) shows KEK data at 
766.2 MeV (895 MeV/c) scaled by $10^{-5}$.} \label{george_qdat}
\end{figure}
\begin{figure}
\caption{$^{12}$C$(\pi^+,\pi^+)$ differential cross sections 
at 400, 500, and 672.5 MeV.  The 400 and 500 MeV data are from this experiment 
and 672.5 MeV (800 MeV/c) data are from BNL.  
Curves are from KDP-RIA calculations.}
\label{cpip_bcc}
\end{figure}
\begin{figure}
\caption{$^{40}$Ca$(\pi^+,\pi^+)$ differential cross sections 
at 400, 500, and 672.5 MeV.  The 400 and 500 MeV data are from this experiment 
and 672.5 MeV (800 MeV/c) data are from BNL.  
Curves are from KDP-RIA calculations.}
\label{capip_bcc}
\end{figure}
\begin{figure}
\caption{$^{40}$Ca$(\pi^-,\pi^-)$ differential cross sections 
at 400, 500, and 672.5 MeV.  The 400 and 500 MeV data are from this experiment 
and 672.5 MeV (800 MeV/c) data are from BNL.  
Curves are from KDP-RIA calculations.}
\label{capim_bcc}
\end{figure}
\begin{figure}
\caption{$^{208}$Pb$(\pi^+,\pi^+)$ differential cross sections 
at 400, 500, and 662.7 MeV.  The 400 and 500 MeV data are from this experiment 
and 662.7 MeV (790 MeV/c) data are from KEK.  
Curves are from KDP-RIA calculations.}
\label{pbpip_bcc}
\end{figure}
\begin{figure}
\caption{$^{208}$Pb$(\pi^-,\pi^-)$ differential cross sections 
at 400, 500, and 662.7 MeV.  The 400 and 500 MeV data are from this experiment 
and 662.7 MeV (790 MeV/c) data are from KEK.  
Curves are from KDP-RIA calculations.}
\label{pbpim_bcc}
\end{figure}

\begin{references}
\bibitem{fnote}Figures showing the KDP-RIA scalar, vector, and effective 
central potentials, tabulations of the calculated and experimental 
small angle cross sections, and figures showing the effects of changing 
the input amplitudes and densities in the KDP-RIA calculations are 
available for all targets and energies from bcc$@$mps.ohio-state.edu or 
kurth$@$mps.ohio-state.edu.
\end{references}
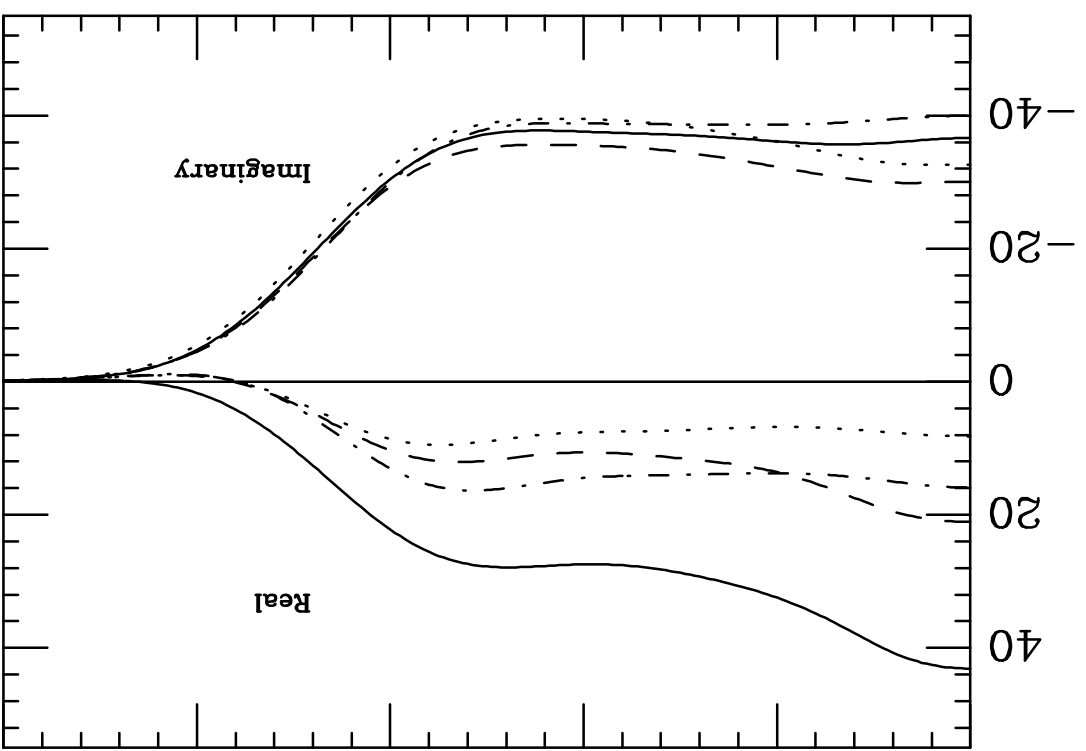
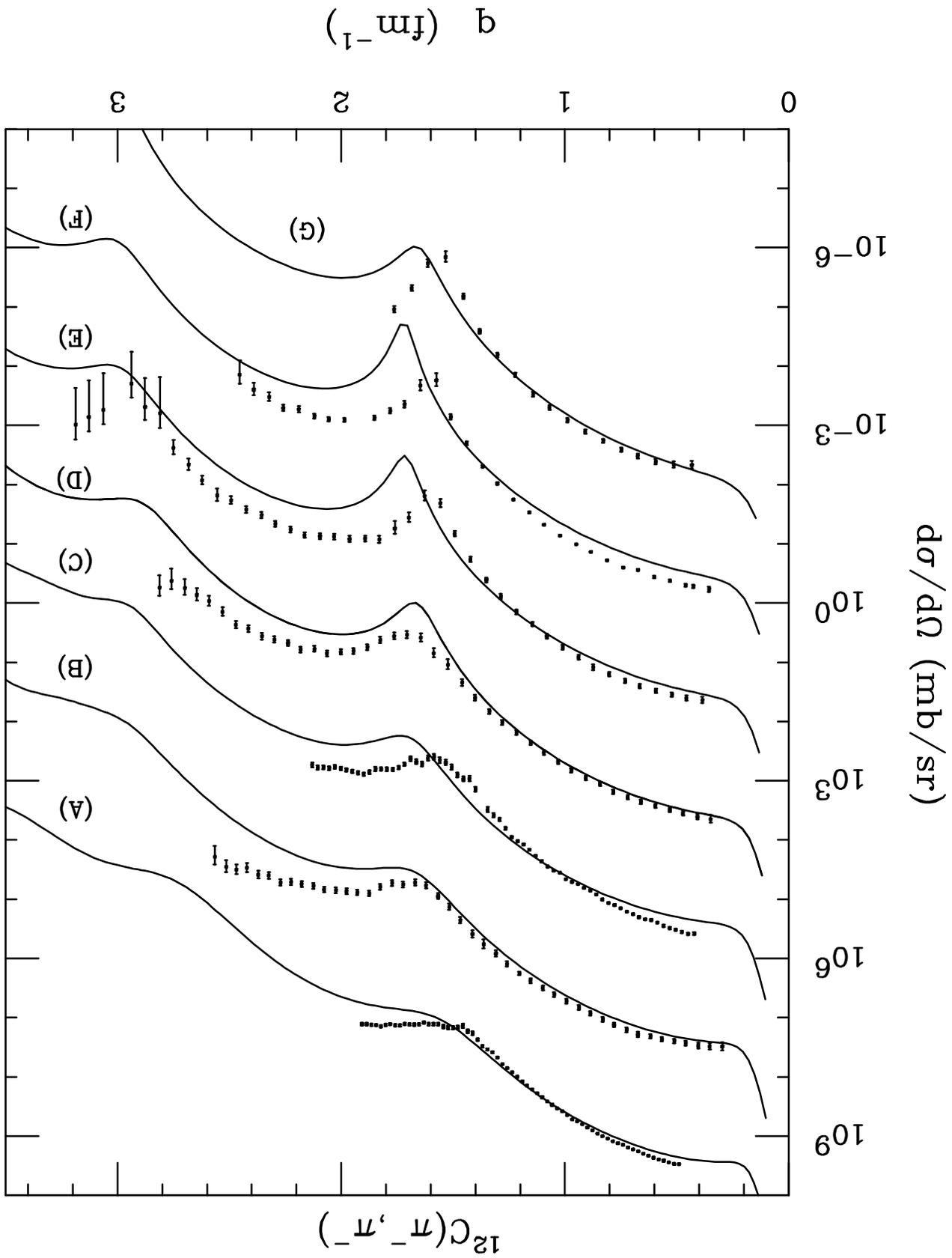
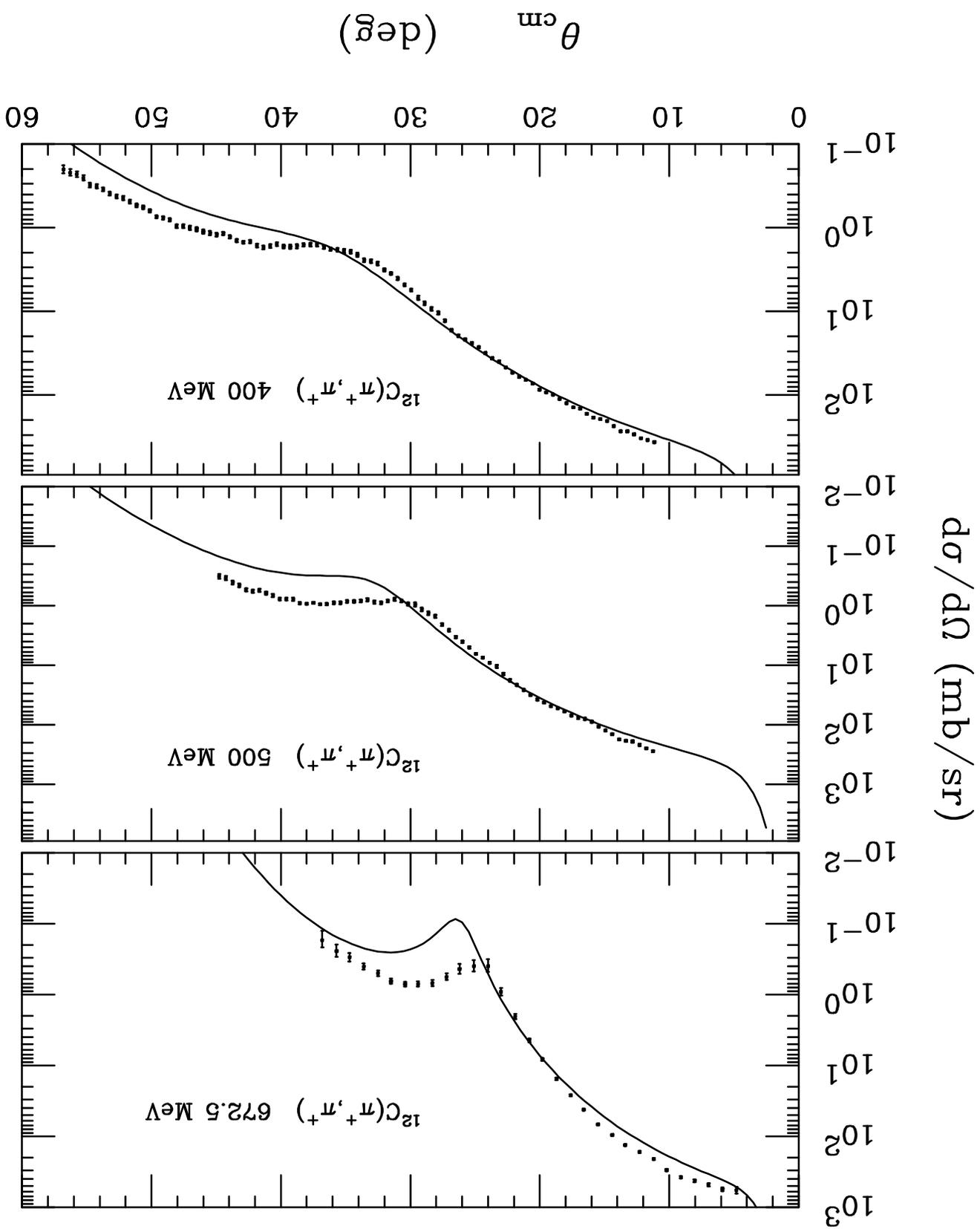
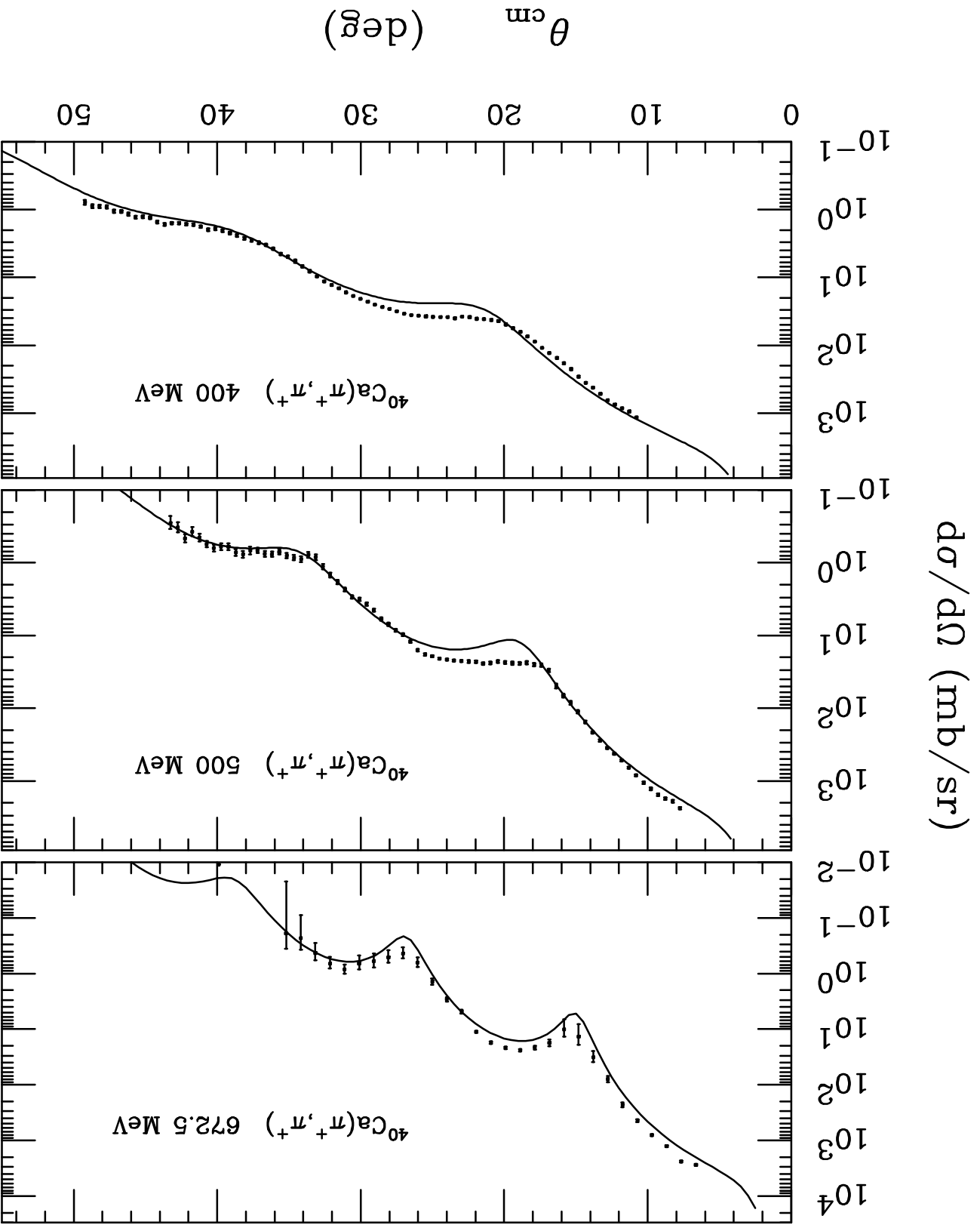
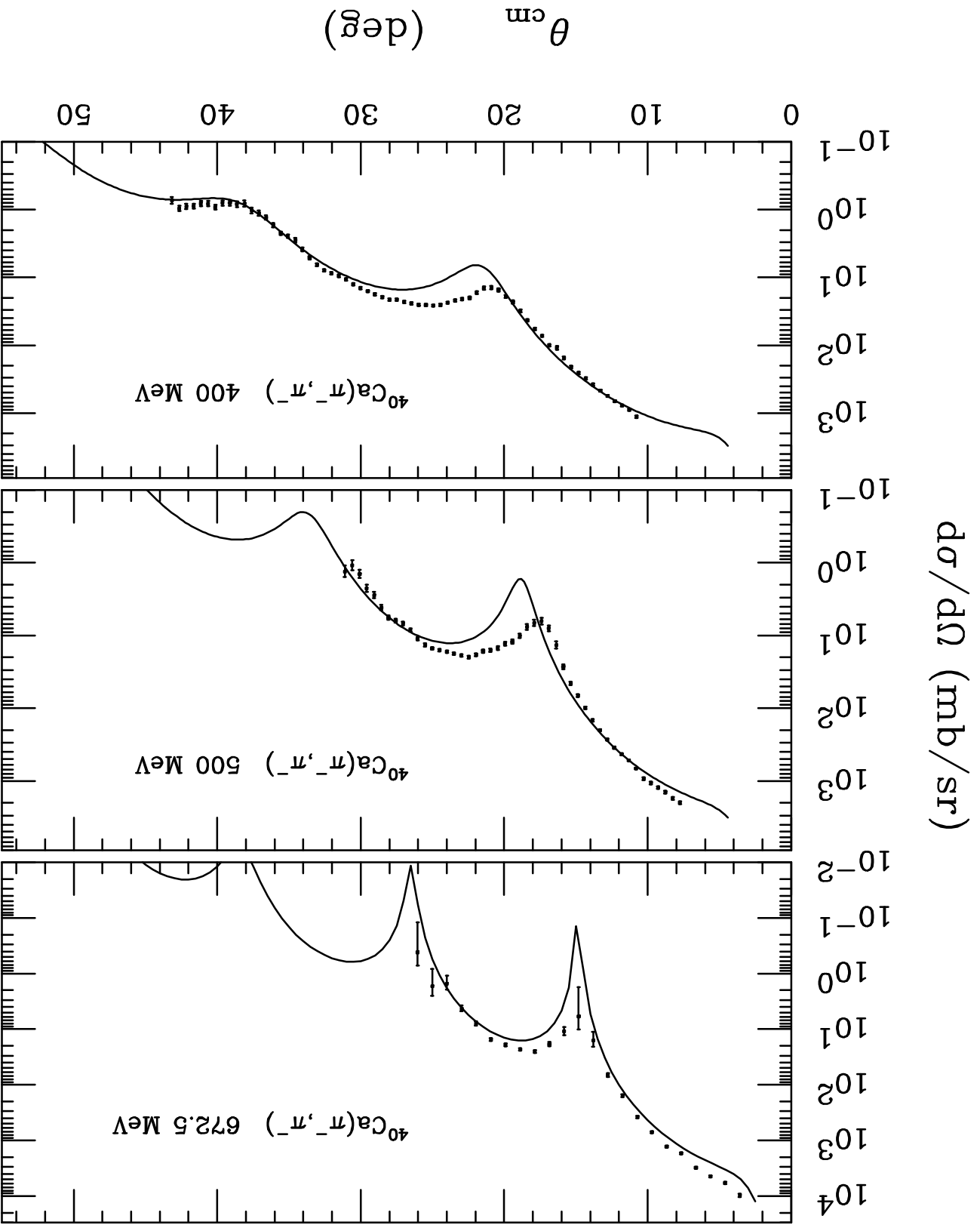
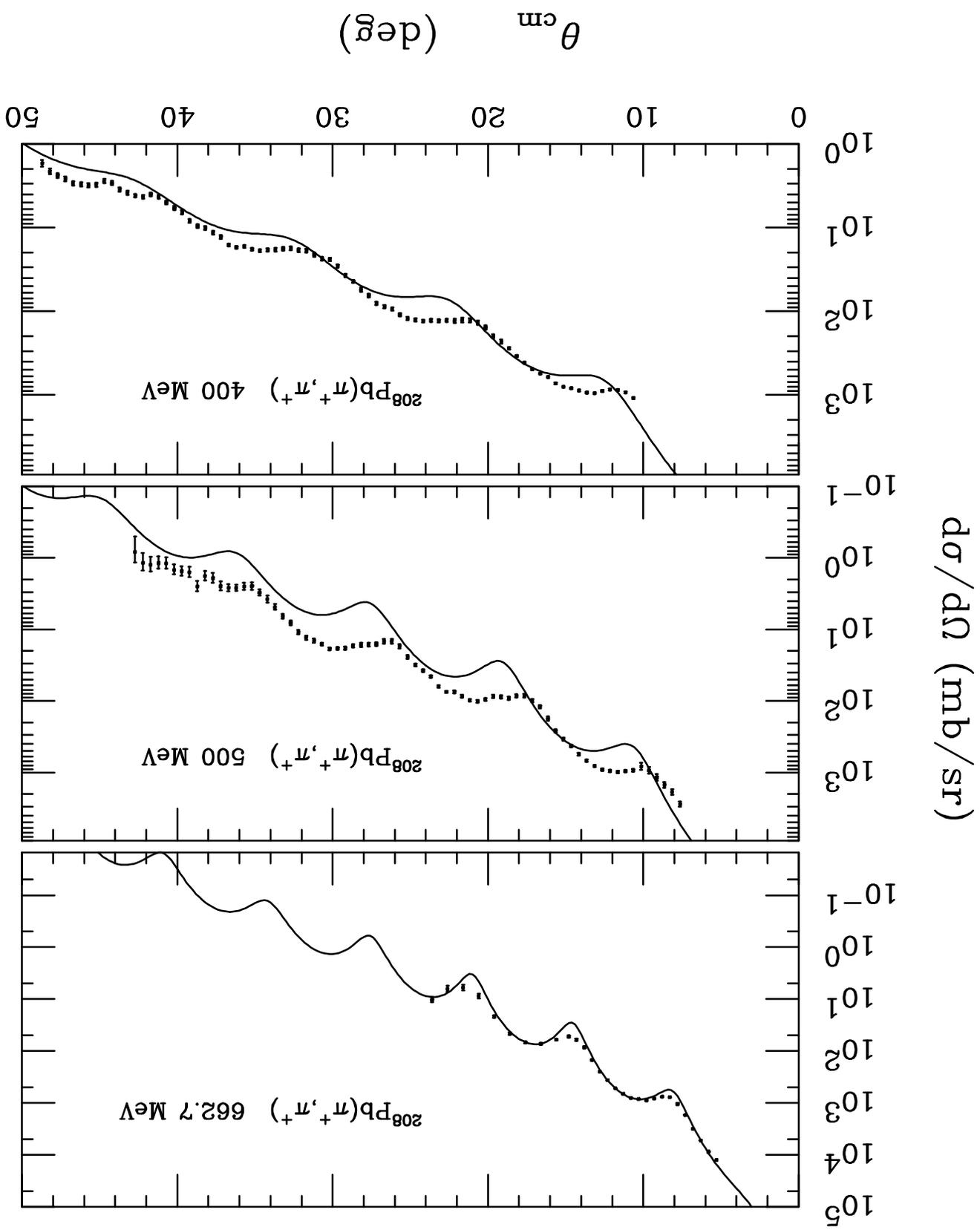
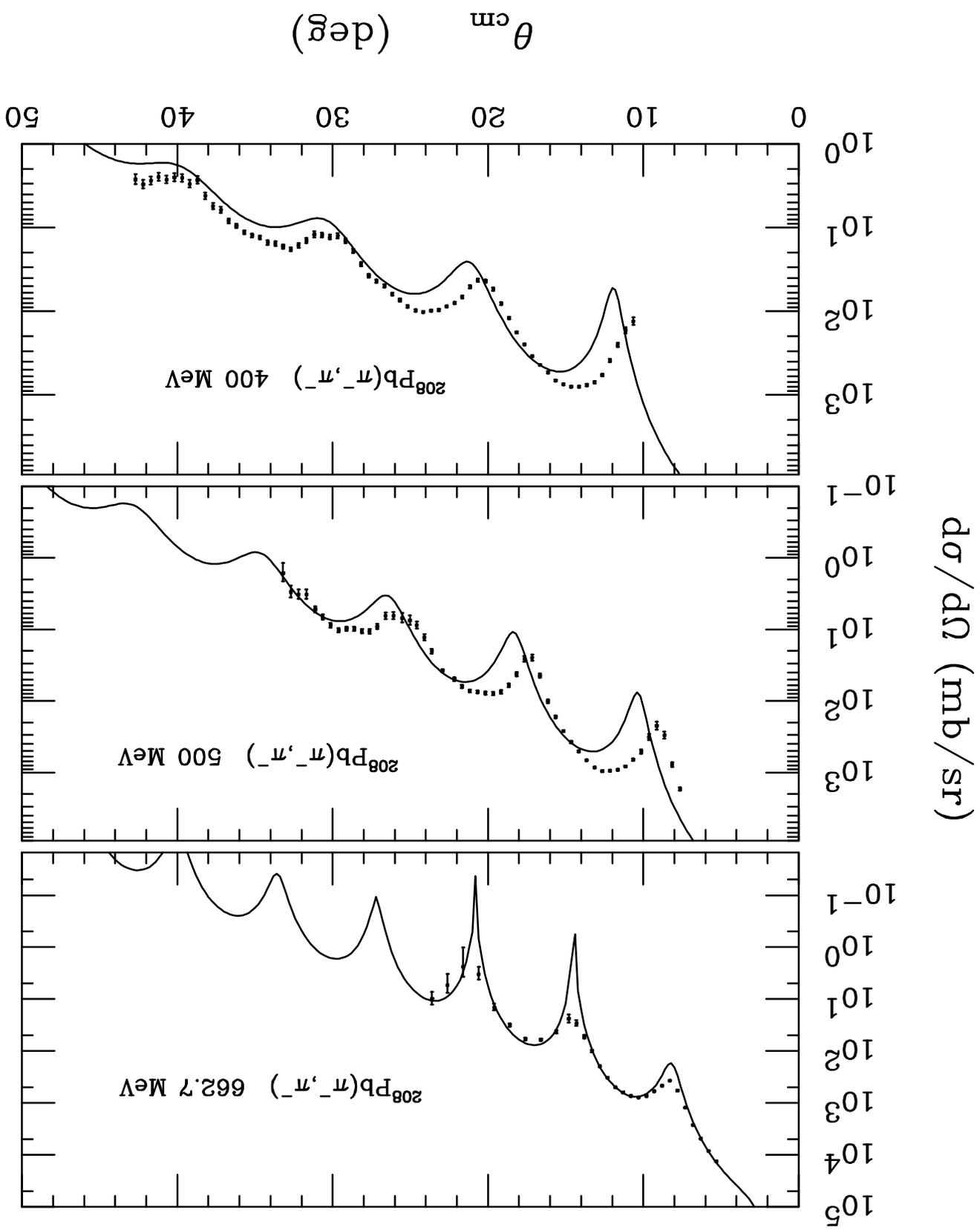
\end{document}